\documentclass[
reprint,
superscriptaddress,
amsmath,amssymb,
aps,prl, 
nofootinbib
]{revtex4-2}

\usepackage[colorlinks=true,allcolors=blue]{hyperref}
\usepackage{graphicx}
\usepackage{orcidlink}
\usepackage{comment}

\newcommand{\pkg}[1]{\texttt{#1}}

\begin{document}

\title{Leakage Mobility in Superconducting Qubits as a Leakage Reduction Unit}
\author{Joan Camps\,\orcidlink{0000-0003-3896-7922}} 
\affiliation{Riverlane, Cambridge, CB2 3BZ, UK} 

\author{Ophelia Crawford\,\orcidlink{0000-0001-7654-1661}} 
\affiliation{Riverlane, Cambridge, CB2 3BZ, UK} 

\author{Gy\"orgy~P.~Geh\'er\,\orcidlink{0000-0003-1499-3229}} 
\affiliation{Riverlane, Cambridge, CB2 3BZ, UK} 

\author{Alexander~V.~Gramolin\,\orcidlink{0000-0001-5436-7375}} 
\email{alexander.gramolin@riverlane.com}
\affiliation{Riverlane, Cambridge, Massachusetts 02142, USA} 

\author{Matthew P. Stafford\,\orcidlink{0000-0001-9938-3854}} 
\affiliation{Riverlane, Cambridge, CB2 3BZ, UK} 
\affiliation{Quantum Engineering Technology Labs, H. H. Wills Physics Laboratory and Department of Electrical \& Electronic Engineering, University of Bristol, BS8 1FD, UK}

\author{Mark Turner\,\orcidlink{0000-0002-5274-957X}} 
\affiliation{Riverlane, Cambridge, CB2 3BZ, UK} 

\date{06 June 2024}

\begin{abstract}
Leakage from the computational subspace is a damaging source of noise that degrades the performance of most qubit types. Unlike other types of noise, leakage cannot be overcome by standard quantum error correction techniques and requires dedicated leakage reduction units. In this work, we study the effects of leakage mobility between superconducting qubits on the performance of a quantum stability experiment, which is a benchmark for fault-tolerant logical computation. Using the Fujitsu Quantum Simulator, we perform full density-matrix simulations of stability experiments implemented on the surface code. We observe improved performance with increased mobility, suggesting leakage mobility can itself act as a leakage reduction unit by naturally moving leakage from data to auxiliary qubits, where it is removed upon reset. We compare the performance of standard error-correction circuits with ``patch wiggling'',  a specific leakage reduction technique where data and auxiliary qubits alternate their roles in each round of error correction. We observe that patch wiggling becomes inefficient with increased leakage mobility, in contrast to the improved performance of standard circuits. These observations suggest that the damage of leakage can be overcome by stimulating leakage mobility between qubits without the need for a dedicated leakage reduction unit.
\end{abstract}
\maketitle

\paragraph{Introduction.}
To achieve practical utility, quantum computers must be able to perform billions of reliable operations. Architectures based on superconducting qubits are currently one of the leading hardware candidates as they benefit from fast gate times and good error rates \cite{nakamura_coherent_1999, arute_quantum_2019, wu_strong_2021, kjaergaard_superconducting_2020, bravyi_future_2022, Li_error_2023}. However, even the most optimistic hardware improvements in error rates will not be enough to reach utility on their own. Quantum error correction (QEC) is required to bridge the gap. In QEC, logical qubits are redundantly encoded into a larger number of noisy physical qubits. Provided the noise suffered by the physical qubits keeps them in the computational subspace and is below a threshold, logical errors can be suppressed to arbitrarily low levels---enabling fault-tolerant computation \cite{aharonov_fault-tolerant_1999}. However, realistic noise also includes other error mechanisms such as leakage, where the qubit leaves the $\{|0\rangle, |1\rangle\}$ computational subspace in an uncontrolled manner. Leakage is particularly damaging as it is long-lived and a leaked qubit can propagate errors to other qubits it interacts with \cite{suchara_leakage_2015, miao_overcoming_2022, acharya_suppressing_2023, wood_quantification_2018, ghosh_understanding_2013}. Standard QEC approaches have no mechanism to deal with leakage. If left unchecked, leakage will eventually corrupt the logical information, even if leakage is orders of magnitude less frequent than other errors~\cite{fowler_coping_2013}. Managing leakage is therefore critical to realizing fault tolerance. 

Leakage can be mitigated with leakage reduction units (LRUs), which remove leakage from the system by regularly returning each qubit to the computational subspace, countering leakage's long-lived nature. When an LRU is included in a QEC scheme, a threshold can be obtained for leakage noise \cite{aliferis_fault-tolerant_2006}. LRUs can broadly be split into two categories: hardware-based, which directly return qubits to the computational space \cite{mcewen_removing_2021, miao_overcoming_2022, battistel_hardware-efficient_2021, marques_all-microwave_2023-1, lacroix_fast_2023}, and circuit-based, which modify the QEC circuit to ensure all qubits are regularly reset---returning them to the computational space if they were leaked \cite{aliferis_fault-tolerant_2006, suchara_leakage_2015}.  Both types of LRU have costs associated with them: hardware-based LRUs often inject more computational noise into the system, whereas circuit-based LRUs require extra qubits and gates. 

Most QEC protocols use two types of physical qubits: auxiliary qubits, which are reset regularly, and data qubits, which cannot be reset without destroying the encoded logical information. In a circuit-based LRU, the QEC circuit is modified so that the roles between auxiliary and data qubits are regularly exchanged, thus allowing every qubit to be reset periodically \cite{ghosh_leakage-resilient_2015, suchara_leakage_2015, brown_handling_2019, brown_leakage_2019, brown_critical_2020, mcewen_relaxing_2023}. In superconducting platforms, leakage can naturally move between qubits, a property known as leakage mobility~\cite{varbanov_leakage_2020}. Therefore, leakage could in principle move from data to auxiliary qubits without the need for a circuit-based LRU. This observation is one of the key insights in this work. We find that, when leakage mobility increases, more leakage is removed from the system and QEC performance improves. That is, leakage mobility itself acts as an LRU. We compare leakage mobility with a recent version of a circuit-based LRU for the surface code~\cite{kitaev_fault-tolerant_2003}, called patch wiggling~\cite{mcewen_relaxing_2023, geher_error-corrected_2023}---the least costly circuit-based LRU for surface codes. We find that patch wiggling is an effective LRU at low leakage mobility, but it is counterproductive at high mobility. Our findings are based on full density-matrix simulations of the stability experiment~\cite{gidney_stability_2022}, a benchmark for QEC performance during fault-tolerant computation. We simulate small instances of the surface code under a realistic noise model for superconducting transmon qubits that includes leakage. 

\paragraph{Surface code.}
The surface code~\cite{kitaev_fault-tolerant_2003} is a current favorable candidate for implementation on superconducting platforms as it only requires a two-dimensional array of qubits with nearest-neighbor connectivity. Performance-wise, it exhibits a high threshold under circuit-level Pauli noise \cite{gidney_stability_2022,raussendorf_topological_2007, fowler_surface_2012}, and can be efficiently decoded using decoders such as minimum-weight perfect matching (MWPM) \cite{dennis_topological_2002, higgott_pymatching_2022} and union-find (UF)~\cite{delfosse_almost-linear_2021}. Furthermore, logical gate implementations are well understood, presenting a clear path to fault-tolerant quantum computation \cite{horsman_surface_2012, chamberland_universal_2021, chamberland_circuit-level_2022, geher_tangling_2023}. An example surface code (for the stability experiment---see later), with which we perform simulations in this paper, is shown in Fig.~\ref{fig:surface_code}(a), laid out on a two-dimensional grid of qubits indicated by the circles.

\begin{figure}[t]
    \centering
    \includegraphics[width = \columnwidth]{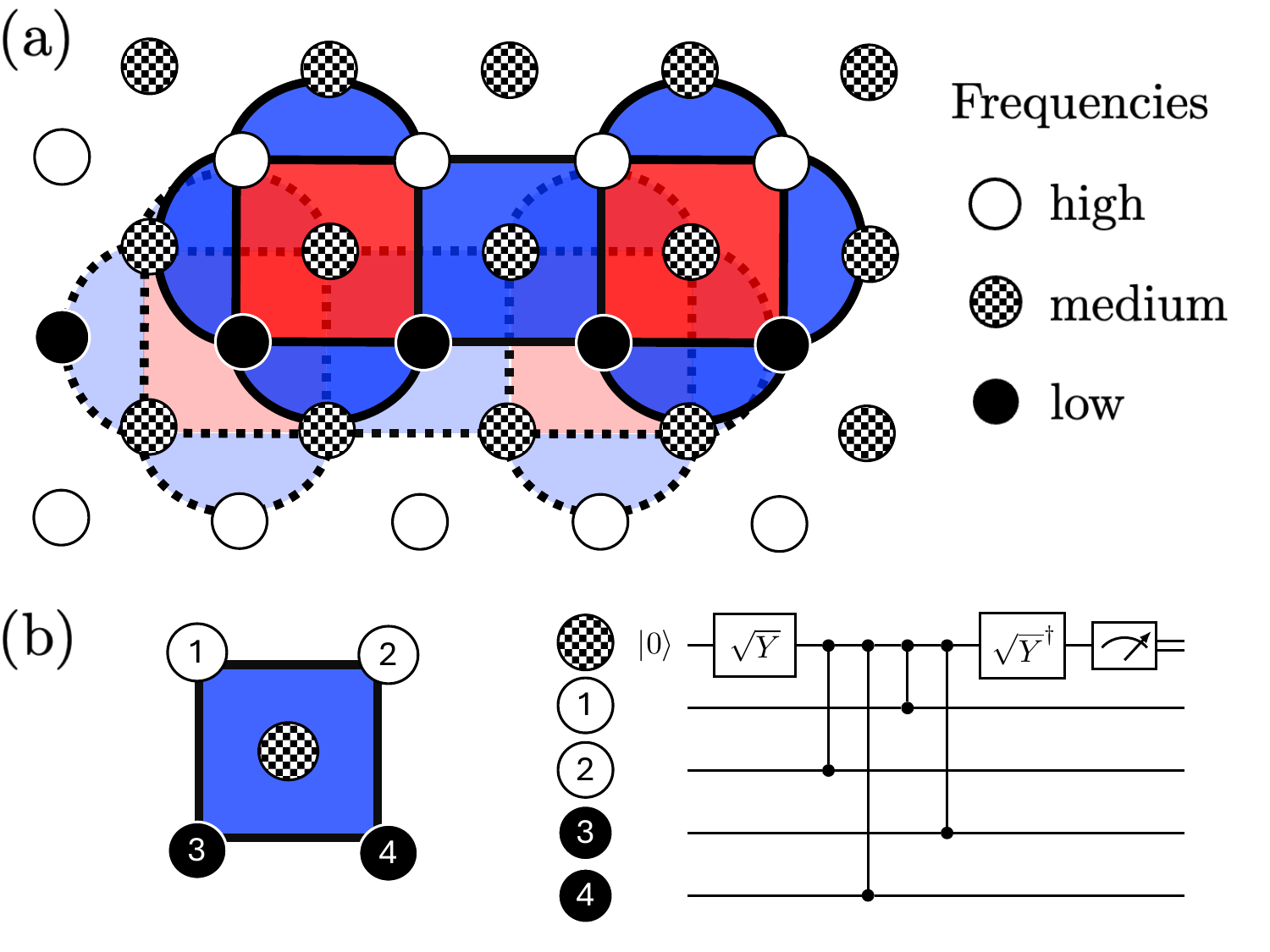}
    \caption{\label{fig:surface_code} (a) The surface code patch for a $4 \times 2$ stability experiment. Data qubits are located on the vertices of the patch with auxiliary qubits on either faces or curved edges. Blue patches indicate $Z$ stabilizers, which are measured with the circuit provided in (b), and red patches correspond to $X$ stabilizers. Qubit frequencies are arranged as shown. During a CZ gate, the higher-frequency qubit of the pair may leak. In a stability experiment, one stabilizer type is over-determined---in this case, the product of all the $Z$ stabilizers is the identity. This is the observable we are protecting. In the patch-wiggling LRU, the location of the code alternates between the dark and faint patches in successive QEC rounds, swapping the qubit roles between data and auxiliary.}
\end{figure}

Error correction is performed by repeatedly measuring a set of operators, called stabilizers. In Fig.~\ref{fig:surface_code}(a), these are indicated by the dark-shaded shapes, with each stabilizer being the product of Pauli operators across the qubits at its corners---$Z$ Paulis for blue shapes and $X$ Paulis for red. Each stabilizer has an associated auxiliary qubit, either in the middle or on the curved edge of each stabilizer in the figure, used to measure the stabilizer. In Fig.~\ref{fig:surface_code}(b) we show an example stabilizer and the circuit used to measure it.

\paragraph{Quantum stability experiment.}
To estimate code performance during logical computation, the simplest benchmarks to perform are quantum memory and stability~\cite{gidney_stability_2022} experiments. Quantum memory is used to evaluate how efficiently a logical qubit can be preserved over time---see, e.g., Ref.~\cite{dennis_topological_2002}. Several recent papers have implemented this experiment in superconducting hardware~\cite{krinner_realizing_2022, acharya_suppressing_2023, ali_reducing_2024} and in simulation with a coherent noise model~\cite{bravyi_correcting_2018, varbanov_leakage_2020, katsuda_simulation_2022, manabe_efficient_2023}. The stability experiment \cite{gidney_stability_2022} is the dual experiment of quantum memory and can be used to evaluate how well logical information can be moved through space. Compared to the memory experiment, a convenient benefit of the stability experiment is that the resilience to errors improves by increasing the number of QEC rounds, rather than the number of qubits---time and space exchange roles. This feature of stability enables exponential suppression of errors in full density-matrix simulations by increasing the runtime without running out of computational resources, in contrast to memory experiments which require increasing the patch size to observe suppression. 

The stability experiment measures the stabilizers of a patch where one type is over-determined. For example, the patch shown in Fig.~\ref{fig:surface_code}(a) has $Z$-type stabilizers that multiply to the identity operator---the logical observable. Benchmarking the stability experiment gives information about how well logical computation can be executed, in particular, a joint logical Pauli measurement (i.e., lattice surgery)~\cite{horsman_surface_2012, chamberland_universal_2021, chamberland_circuit-level_2022}, which is a key ingredient of Pauli-based computation~\cite{bravyi_trading_2016}.

\paragraph{Simulation details.}
We implemented our simulations using the \pkg{Python} programming language and Qulacs---a fast classical simulator of quantum circuits that supports custom multi-qubit gates~\cite{suzuki_Qulacs_2021}. The simulations were run on the Fujitsu Quantum Simulator, which is a computer cluster comprised of 512 Fujitsu A64FX processors, each featuring 48 cores and 32~GB of random-access memory~\cite{imamura_mpiQulacs_2022}. 

To capture the possibility of leakage, we use qutrits---three-level systems, each having an additional, non-computational state $|2\rangle$. Since Qulacs does not natively support qutrits, we model each qutrit using two qubits in the simulator. The qutrit states $|0\rangle$, $|1\rangle$, and $|2\rangle$ can be encoded as the two-qubit states $|\overline{00}\rangle$, $|\overline{01}\rangle$, and $|\overline{10}\rangle$, respectively---where qubits in the simulator are denoted with a bar. Similarly, the nine two-qutrit states can be represented using four qubits: $|00\rangle \rightarrow |\overline{0000}\rangle$, $|01\rangle \rightarrow |\overline{0001}\rangle$, $|02\rangle \rightarrow |\overline{0010}\rangle$, $\ldots\,$, $|22\rangle \rightarrow |\overline{1010}\rangle$. In this way, single- and two-qutrit gates can be modeled as two- and four-qubit gates, respectively.

We adapt the noise model of Ref.~\cite{varbanov_leakage_2020} that describes a flux-tunable transmon architecture~\cite{krantz_guide_2019, blais_circuit_2021}, where qubits have three frequency types and are arranged as shown in Fig.~\ref{fig:surface_code} \cite{versluis_scalable_2017}. Note that other transmon architectures might have different~\cite{krinner_realizing_2022} or, in the case of tunable couplers, less ordered~\cite{miao_overcoming_2022} frequency arrangements.

In this work, we use the $\sqrt{Y}$ and $\sqrt{Y}^{\dagger}$ single-qutrit gates and the CZ two-qutrit gate---see Fig.~\ref{fig:surface_code}(b). Since CZ gates represent the dominant source of leakage noise, we follow Ref.~\cite{varbanov_leakage_2020} and assume that single-qutrit gates do not induce leakage and act on $|2\rangle$ as the identity. The corresponding two-qubit gates are defined in Qulacs by their $4 \times 4$ unitary matrices. We also use two other types of single-qutrit operations: measurement and reset. Measurements are implemented in Qulacs as projections onto the \{$|0\rangle$, $|1\rangle$, $|2\rangle$\} basis. Each auxiliary qutrit is measured at the end of a QEC round and then reset to $|0\rangle$.

\begin{figure}
    \centering
    \includegraphics[width = \columnwidth]{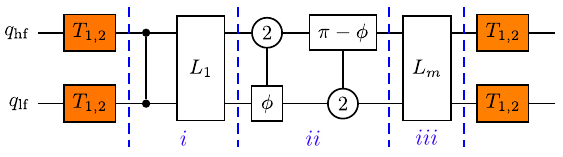}
    \caption{\label{fig:CZ_model}Model for the noisy two-qutrit CZ gate~\cite{varbanov_leakage_2020}. The higher- and lower-frequency qutrits are denoted as $q_{\text{hf}}$ and $q_{\text{lf}}$, respectively. The single-qutrit channels $T_{1,2}$ represent the relaxation and dephasing processes. The two-qutrit operations correspond to Eqs.~(\ref{eq:1})--(\ref{eq:4}) and include the following components: (\emph{i}) an ideal CZ gate followed by a gate describing leakage with probability $L_1$, (\emph{ii}) two single-qutrit rotations, by phases $\phi$ and ($\pi - \phi$), conditional on the other qutrit being in $|2\rangle$, and (\emph{iii}) a gate describing leakage mobility with probability $L_m$.}
\end{figure}

In our simulations, we account for qutrit relaxation and dephasing via amplitude- and phase-damping processes, respectively. We define the corresponding $T_{1,2}(t)$ channel in Qulacs using Kraus operators obtained by numerically solving the Lindblad master equation~\cite{lindblad_generators_1976} for the time evolution of a three-level transmon (see Supplemental Material and Ref.~\cite{varbanov_leakage_2020} for details). We assume that each of the four possible operations---single-qutrit gate, two-qutrit gate, measurement, and reset---has a specific time duration. At each point in time, the qutrit can either undergo an operation or be idle (when waiting for an operation on another qutrit to be completed). For qutrit idling of duration $\tau$, we apply the $T_{1,2}(\tau)$ channel. For each single-qutrit operation, such as $\sqrt{Y}$, we apply the $T_{1,2}$ channel symmetrically around it: $T_{1,2}(\tau/2) \sqrt{Y} \, T_{1,2}(\tau/2)$. Similarly, we place each CZ gate of duration $\tau$ between four $T_{1,2}(\tau/2)$ channels, as shown in Fig.~\ref{fig:CZ_model}. All parameters corresponding to the above processes are summarized in Table~\ref{tab:parameters}.

For CZ gates, we adapt the noise model of Ref.~\cite{varbanov_leakage_2020}, which is illustrated in Fig.~\ref{fig:CZ_model}. This model assumes flux-tunable superconducting transmon qubits with static coupling and the ``11--02'' implementation of the CZ gate~\cite{blais_circuit_2021, rol_fast_2019}. In this implementation, the qubit $q_{\text{hf}}$ has a higher ``sweet-spot'' frequency and is flux-tunable, while the qubit $q_{\text{lf}}$ has a lower frequency that remains static. To realize the CZ gate, a flux pulse is applied to $q_{\text{hf}}$ to briefly bring its frequency down to the avoided crossing between $|11\rangle$ and $|02\rangle$ (where the notation $|q_{\text{lf}} \, q_{\text{hf}}\rangle$ is assumed). For this reason, the higher-frequency qubit is susceptible to leakage, while the lower-frequency one has a negligible leakage probability.

The CZ gate with leakage (see Fig.~\ref{fig:CZ_model}(\emph{i})) is modeled as an exchange between states $|11\rangle$ and $|02\rangle$, such that
\begin{align}
|11\rangle &\rightarrow -\sqrt{1 - 4L_1} \, |11\rangle + 2\sqrt{L_1} \, |02\rangle, \label{eq:1} \\
|02\rangle &\rightarrow -\sqrt{1 - 4L_1} \, |02\rangle - 2\sqrt{L_1} \, |11\rangle,
\end{align}
where $0 \le L_1 \le 0.25$ is the leakage probability. The ideal CZ gate is included in Eq.~\eqref{eq:1} in the absence of leakage ($L_1 = 0$), where it becomes $|11\rangle \rightarrow -|11\rangle$. 

When one of the qubits involved in the CZ gate is leaked we see two effects. Firstly, a leakage-conditional phase $\phi$\footnote{Our notation differs from that of Ref.~\cite{varbanov_leakage_2020}, where $\phi_{\text{stat}}^{\mathcal{L}} = \phi$ and $\phi_{\text{flux}}^{\mathcal{L}} = \pi - \phi$.}, conditioned on one of the qubits being in the $|2\rangle$ state, is applied---see Fig.~\ref{fig:CZ_model}(\emph{ii}). The applied phase is dependent upon whether the lower- or higher-frequency qubit is leaked. This phase is the subsequent computational noise introduced to the system when a leaked qubit undergoes further CZ gates. We fix $\phi=\pi/10$ across all gates in our simulation. The second effect is leakage mobility between the two qubits, denoted $L_m$ in Fig.~\ref{fig:CZ_model}(\emph{iii}), which is fuelled by accidental near crossings during the 11--02 gate, and is modeled as an exchange between $|21\rangle$ and $|12\rangle$. The combination of these effects is given by
\begin{align}
|21\rangle &\rightarrow -\sqrt{1 - 4L_m} e^{+i\phi} \, |21\rangle + 2\sqrt{L_m} \, |12\rangle, \label{eq:3} \\
|12\rangle &\rightarrow -\sqrt{1 - 4L_m} e^{-i\phi} \, |12\rangle - 2\sqrt{L_m} \, |21\rangle, \label{eq:4}
\end{align}
where $0 \le L_m \le 0.25$ is the leakage mobility parameter. We vary $L_1$ and $L_m$ to cover values reported in Refs. \cite{varbanov_leakage_2020, miao_overcoming_2022}, as shown in Table~\ref{tab:parameters}. Equations~(\ref{eq:1})--(\ref{eq:4}) fully determine the $9 \times 9$ unitary matrix describing our two-qutrit CZ gate (see Supplementary Material). The corresponding four-qubit gate is implemented in Qulacs by specifying its $16 \times 16$ unitary matrix.

\begin{table}
\caption{\label{tab:parameters}Numerical parameters used in the simulation.}
\begin{ruledtabular}
\begin{tabular}{lr}
Parameter & Numerical value \\
\hline
Relaxation time $T_1$ & $30~\mu\text{s}$ \\
Dephasing time $T_2 = T_{\phi}/2$ & $30~\mu\text{s}$ \\
$\sqrt{Y}$, $\sqrt{Y}^{\dagger}$ gate duration & 20~ns \\
CZ gate duration & 40~ns \\
Measurement duration & 600~ns \\
Reset duration & 500~ns \\
Conditional phase $\phi$ & $\pi / 10$ \\
\multicolumn{2}{c}{Variable parameters:} \\
Leakage $L_1$ & 0.1\%, 0.5\% \\
Leakage mobility $L_m$ & 0.5\%, 2.5\%, 6\%, 9\%, 12.5\% \\
\end{tabular}
\end{ruledtabular}
\end{table}

To reduce the number of qubits we need to simulate in Qulacs, we measure each stabilizer sequentially in simulation, so that auxiliary qubits can be recycled~\cite{obrien_density-matrix_2017, katsuda_simulation_2022}. By applying relaxation and dephasing noise corresponding to that in the circuit without recycling, the overall behavior of the system is unchanged, and we effectively simulate simultaneous syndrome extraction for all the stabilizers. In this way, we reduce the effective count of auxiliary qutrits to just one. Recall that we need two Qulacs qubits to simulate a single qutrit. For circuits with/without wiggling, recycling reduces the simulation qubit count from 44/34 to 18 for the $4 \times 2$ patch of Fig.~\ref{fig:surface_code}.

\paragraph{Decoding.}
We decode our simulations with the MWPM decoder \cite{dennis_topological_2002, higgott_pymatching_2022}, in which error mechanisms are represented by edges of a graph and the combinations of measurement outcomes (which change value due to a particular error mechanism) are represented by the vertices at either end of the edge. The graph is weighted with each edge weight representing the probability of the corresponding error. The decoder finds a matching with minimal weight---the most likely set of errors explaining the observed measurement outcomes.
 
We employ an experimentally-derived decoding graph \cite{spitz_adaptive_2018, chen_calibrated_2022, chen_exponential_2021, acharya_suppressing_2023}. This is constructed in three steps: first, we use Stim \cite{gidney_stim_2021} to generate a decoding graph based on a simplified error model without leakage but with approximated Pauli errors on resets, measurements, idling, and single-qubit and two-qubit gates; second, we forego the weights of this graph; third, we fit new weights to this decoding graph using half of the shots of the experiment---the training data set. We then discard the training data and decode the remaining $n_{\textrm{shots}}$ shots using the updated weights. We repeat this process for every error model in our simulations. As the decoder is not ``leakage-aware''---i.e., it does not recognize a ``2'' measurement result---we assume that measurements can only return ``0'' or ``1'' and treat the outcome ``2'' as if it were ``1''.

A logical error occurs when the observed logical outcome (in our case, the product of all the $Z$ stabilizers in Fig.~\ref{fig:surface_code}) differs from the outcome predicted by the decoder.  The frequency of this discrepancy is the probability of logical failure $P_{\textrm{fail}}=n_{\textrm{failures}}/n_{\textrm{shots}}$, and its error bar is $\delta P_{\textrm{fail}} = \sqrt{P_{\textrm{fail}}(1-P_{\textrm{fail}})/n_{\textrm{shots}}}$.

\paragraph{Results.}
Our main results are summarized in 
Fig.~\ref{fig:intersection_4x2}, where we plot the error suppression rate $\Lambda_s$ as a function of the mobility parameter $L_m$. We define $\Lambda_s$ of the stability experiment as the base of the exponential suppression of the logical error with the number of rounds. That is, for a number of rounds $n_r$, we compute the probability of logical failure $P_{\textrm{fail}}$, and then obtain $P_{\textrm{SPAM}}$ and $\Lambda_s$ in
\begin{equation}
    P_{\textrm{fail}} = P_{\textrm{SPAM}}(\Lambda_s)^{-n_r/2}\,,
\end{equation}
through a linear fit in a $(n_r,\,\,  \log(P_{\textrm{fail}}))$ plot. For the data in Fig.~\ref{fig:intersection_4x2}, we perform simulations with $n_r \in\{3, 5, 7, 9\}$, each with a number of shots between $9 \times 10^4$ and $3 \times 10^5$. This error suppression rate $\Lambda_s$ in stability experiments is analogous to $\Lambda$ in quantum memory experiments (see, e.g., Ref.~\cite{acharya_suppressing_2023}). Being below threshold corresponds to $\Lambda_s>1$, and higher $\Lambda_s$ indicates more error suppression.

\begin{figure}
    \centering
    \includegraphics[width = \columnwidth]{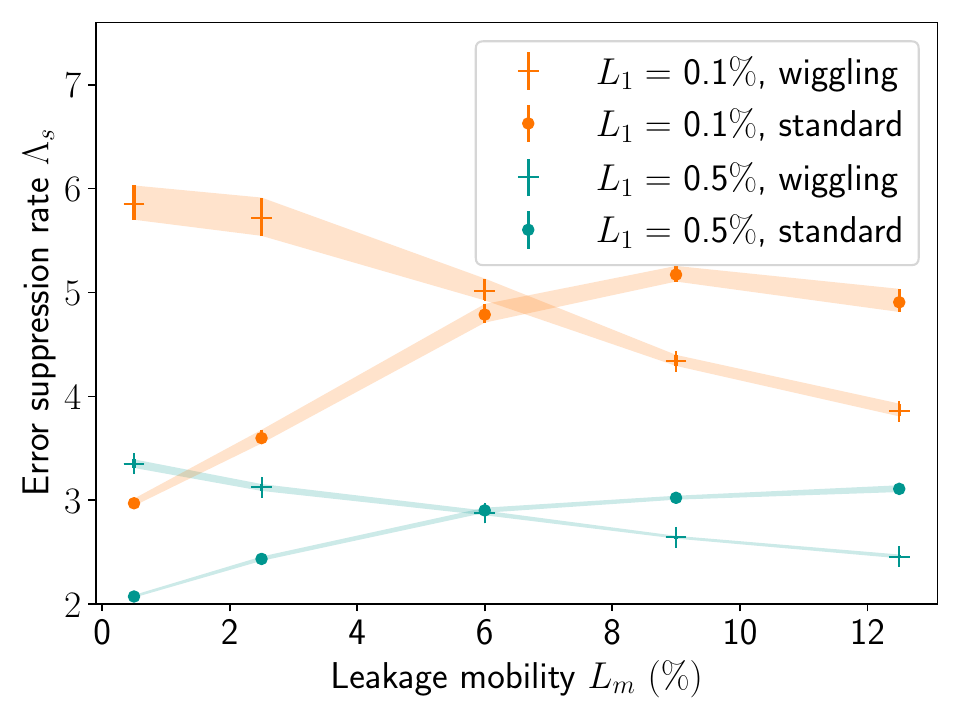}
    \caption{Wiggled (crosses) and standard, non-wiggled (dots) error suppression rates as a function of the leakage mobility parameter $L_m$. Orange data are for low levels of leakage $L_1=0.1\%$; green data are for $L_1 = 0.5\%$. Wiggled circuits achieve higher error suppression at low mobility, whereas standard circuits are better at high mobility.}
    \label{fig:intersection_4x2}
\end{figure}

For standard circuits (dots), logical performance improves with increasing $L_m$, indicating that mobility itself acts as an LRU. This is our main result. If leakage moves efficiently, it can transfer from data to auxiliary qubits during a round of syndrome extraction, after which it is removed during reset (see Supplemental Material for more details). In contrast, the logical performance of the wiggled circuits (crosses) degrades with increasing $L_m$, suggesting that wiggling becomes counterproductive at high $L_m$.

Choosing whether to implement the patch-wiggling LRU or not is dependent upon the value of $L_m$. At low mobility values, wiggling greatly improves performance compared to the standard circuit. For instance, the suppression factor $\Lambda_s$ of the wiggled circuit is double that of the standard circuit at $L_1=0.1\%$, $L_m=0.5\%$. However, we find that at large values of $L_m$, the standard, non-wiggled circuits outperform the wiggled ones. The crossover is at around $L_m \sim 6\%$, both at high ($0.5\%$) and low ($0.1\%$) values of the leakage parameter $L_1$. 

\paragraph{Discussion and outlook.}
We have performed the first full density-matrix simulations of the stability experiment.  We have considered a noise model with leakage and explored the role of leakage mobility as an LRU. We have seen that increasing leakage mobility can act as an LRU and boost error suppression. Compared to other LRUs, such as patch wiggling, leakage mobility is cheaper as it has no overhead in qubits or gates. A natural question is whether the effects we have seen are a feature of the patch we have simulated (Fig.~\ref{fig:surface_code}). To address this, we have performed stability experiment simulations on a $3 \times 3$ surface code patch (see Supplemental Material), and we have observed the same qualitative effects.

Although our simulations are for a qubit frequency arrangement and noise models inspired by the tunable transmon model of Ref.~\cite{varbanov_leakage_2020}, we expect that our observations apply more generally, including for tunable coupler architectures~\cite{miao_overcoming_2022}, where the effective frequency arrangement may be less ordered. While we have focused on the stability experiment for the surface code, we expect that analogous conclusions regarding leakage mobility as an LRU will hold for quantum memory experiments and in other QEC codes---mobility facilitates the transfer of leakage from data to auxiliary qubits (where it is removed upon reset) regardless of the structure of the code.  We also expect that the loss of effectiveness of patch wiggling as an LRU in the presence of high coherent mobility holds in other QEC benchmarks. We hope that future work will confirm these expectations. Finally, a natural suggestion for further study is to assess the feasibility of designing superconducting qubits with engineered high leakage mobility as a natural mechanism to deal with leakage.

\paragraph{Acknowledgements.}
The authors thank Earl Campbell, Marcin Jastrzebski, Hari Krovi, Elisha Siddiqui Matekole, Setiawan, and Jacob Taylor for many insightful discussions and their help during various phases of the project. We are also grateful to Earl Campbell and Setiawan for their comments on the manuscript. This work would not have been possible without the support from Fujitsu Limited, which organized the Quantum Simulator Challenge and provided us with generous access to their computer cluster.

\paragraph{Author contributions.}
J.C. and A.V.G. supervised the project. O.C. and G.P.G. designed the quantum circuits and generated the corresponding Stim files. M.P.S. reviewed the literature and evaluated different approaches to modeling leakage. A.V.G. performed the simulations in Qulacs with support from M.T.. J.C. analyzed the simulation data. All authors contributed to conceiving and shaping the project, interpreting its outcomes, and writing the manuscript.

\clearpage
\appendix

\setcounter{equation}{0}
\setcounter{figure}{0}
\renewcommand{\theequation}{S\arabic{equation}}
\renewcommand{\thefigure}{S\arabic{figure}}

\begin{titlepage}
\centering
\Large
Supplemental Material for\\ ``Leakage Mobility in Superconducting Qubits as a Leakage Reduction Unit''
\end{titlepage}

\onecolumngrid

\section{Relaxation and dephasing channels}

We account for qutrit decoherence by applying the amplitude-damping (``relaxation'') and phase-damping (``dephasing'') noise channels~\cite{varbanov_leakage_2020, nielsen_quantum_2010}. In Qulacs, custom gates/channels are implemented as superoperator ``Qulacs gates" that are defined by the corresponding unitary matrices or Kraus operators. To obtain the Kraus operators needed for the relaxation and dephasing channels, we first numerically solve the Lindblad master equation for a density matrix~$\rho$,
\begin{equation}
\frac{d\rho}{dt} = -i[H, \, \rho] + \sum_j L_j \rho L_j^{\dagger} - \frac{1}{2} \left\{L_j^{\dagger} L_j, \, \rho\right\}, \label{eq:Lindblad_equation}
\end{equation}
with the transmon qubit Hamiltonian
\begin{equation}
H = \omega a^{\dagger} a + \frac{\alpha}{2}(a^{\dagger})^2 a^2,
\end{equation}
where $a^{\dag}$ and $a$ are the creation and annihilation operators, $\omega$ and $\alpha$ are qubit's resonance frequency and anharmonicity, and $L_j$ are the Lindblad operators~\cite{varbanov_leakage_2020}. The amplitude-damping (relaxation) process is modeled by the Lindblad operator
\begin{equation}
L_{\text{amp}} = \sqrt{\frac{1}{T_1}} a = \sqrt{\frac{1}{T_1}}
\begin{pmatrix}
0 & 1 & 0 \\
0 & 0 & \sqrt{2} \\
0 & 0 & 0
\end{pmatrix},
\label{eq:T1}\end{equation}
where $T_1 = 30~\mu\text{s}$ is the relaxation time for the $|1\rangle \rightarrow |0\rangle$ transition. Phase damping (dephasing) is described by the Lindblad operators~\cite{varbanov_leakage_2020}
\begin{equation}
L_{\text{deph},1} = \sqrt{\frac{4}{9 T_2}}
\begin{pmatrix}
1 & 0 & 0 \\
0 & 0 & 0 \\
0 & 0 & -1
\end{pmatrix}, \quad
L_{\text{deph},2} = \sqrt{\frac{1}{9 T_2}}
\begin{pmatrix}
1 & 0 & 0 \\
0 & -1 & 0 \\
0 & 0 & 0
\end{pmatrix}, \quad
L_{\text{deph},3} = \sqrt{\frac{1}{9 T_2}}
\begin{pmatrix}
0 & 0 & 0 \\
0 & 1 & 0 \\
0 & 0 & -1
\end{pmatrix},
\label{eq:T2}\end{equation}
where $T_2 = T_{\phi}/2 = 30~\mu\text{s}$ is the dephasing time.

We solve the master Eq.~\eqref{eq:Lindblad_equation} numerically using \pkg{QuTiP} and obtain a Liouvillian superoperator~$\mathcal{L}$, which is a $9\times 9$ matrix describing the overall effect of our decoherence channel. The corresponding time-dependent superoperator $\mathcal{E}(t)$ can be found by matrix exponentiating~$\mathcal{L}$ to give
\begin{equation}
\mathcal{E}(t) = \exp{(t \mathcal{L})},
\end{equation}
where $t$ is the idle time for which the decoherence occurs. Finally, we use \pkg{QuTiP} to convert the $9\times 9$ superoperator $\mathcal{E}$ for a specific idle time~$t$ into a set of $3\times 3$ Kraus operators~$K_i$ such that~\cite{nielsen_quantum_2010}
\begin{equation}
\mathcal{E}(\rho) = \sum\limits_i K_i \rho K_i^{\dag}, \qquad \sum\limits_i K_i^{\dag} K_i = I.
\label{eq:E(rho)}\end{equation}

Using the above approach, we can specify in Qulacs a custom decoherence gate $T_{1,2}(t)$, corresponding to the $\mathcal{E}(\rho)$ channel of Eq.~\eqref{eq:E(rho)} associated to the Lindlad operators of Eqs.~\eqref{eq:T1} and \eqref{eq:T2}, acting for a time interval~$t$. For a single-qutrit gate $U_{\text{gate}}$ of duration $t_{\text{gate}}$, decoherence is accounted for by applying the symmetric gate sequence $T_{1,2}(t_{\text{gate}}/2) \, U_{\text{gate}} \, T_{1,2}(t_{\text{gate}}/2)$, that should be understood as concatenation of Qulacs gates. For qutrit idling of duration $t_{\text{idle}}$, we apply $T_{1,2}(t_{\text{idle}})$.

\section{Unitary matrices for the single- and two-qutrit gates}

For convenience, we provide here explicit unitary matrices for the single- and two-qutrit gates that we used in our simulations. The single-qutrit gates $\sqrt{Y}$ and $\sqrt{Y}^{\dagger}$ are defined by the $3 \times 3$ matrices
\begin{equation}
\sqrt{Y} =
\begin{bmatrix}
\frac{1}{2}(1+i) & -\frac{1}{2}(1+i) & 0 \\
\frac{1}{2}(1+i) & \frac{1}{2}(1+i) & 0 \\
0 & 0 & 1
\end{bmatrix}, \qquad
\sqrt{Y}^{\dag} =
\begin{bmatrix}
\frac{1}{2}(1-i) & -\frac{1}{2}(1-i) & 0 \\
\frac{1}{2}(1-i) & \frac{1}{2}(1-i) & 0 \\
0 & 0 & 1
\end{bmatrix}.
\end{equation}

The two-qutrit CZ gate is defined by the $9 \times 9$ unitary matrix
\begin{equation}
U_{\text{CZ}} =
\begin{bmatrix}
1 & & & & & & & & \\
& 1 & & & & & & & \\
& & -\sqrt{1-4L_1} & & 2\sqrt{L_1} & & & & \\
& & & 1 & & & & & \\
& & -2\sqrt{L_1} & & -\sqrt{1-4L_1} & & & & \\
& & & & & -\sqrt{1-4L_m} \, e^{-i\phi} & & 2\sqrt{L_m} & \\
& & & & & & 1 & & \\
& & & & & -2\sqrt{L_m} & & -\sqrt{1-4L_m} \, e^{i\phi} & \\
& & & & & & & & 1
\end{bmatrix}, \label{eq:U_CZ}
\end{equation}
where the parameters $L_1$, $L_m$, and $\phi$ are described in the main text. Note that we have omitted zero entries in matrix~(\ref{eq:U_CZ}) for clarity.

\section{Additional data for the $4\times 2$ experiment}\label{sec:supp4x2}
\begin{figure}[h!]
    \centering
    \includegraphics[width = \textwidth]{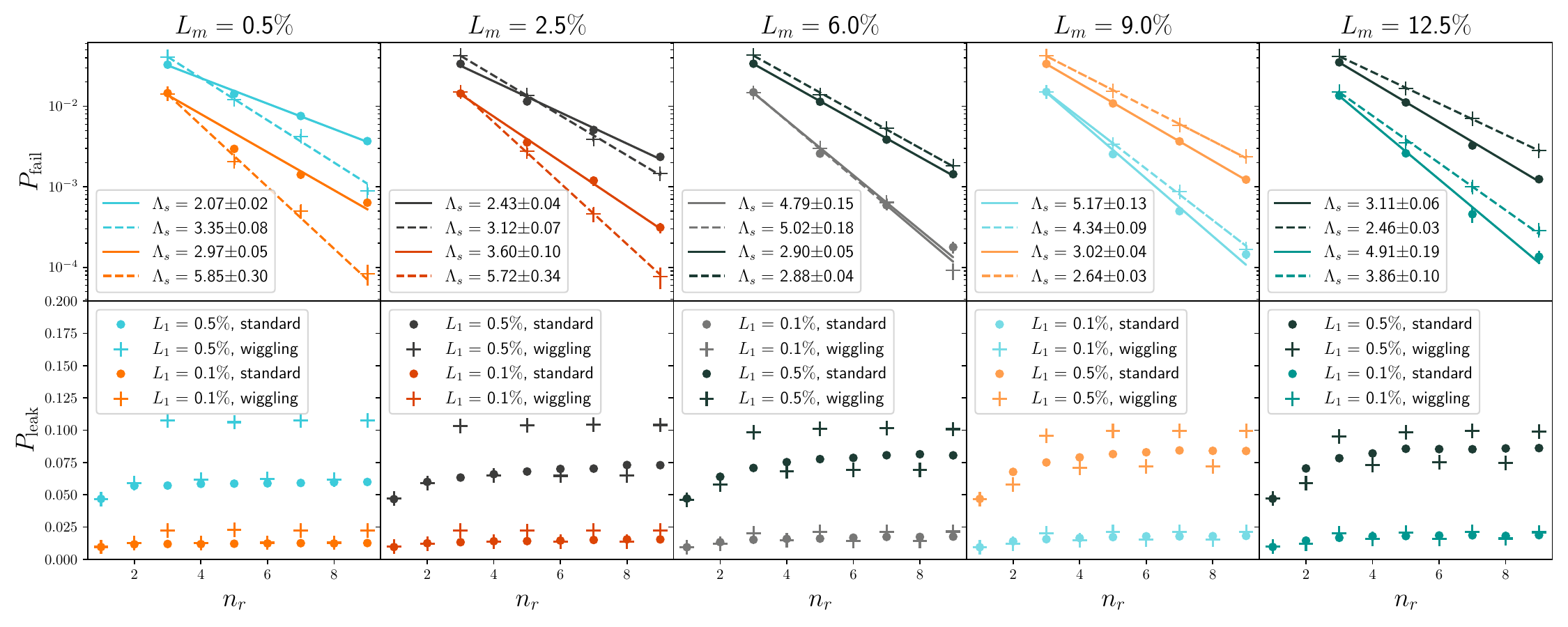}
    \caption{Additional results for the $4\times 2$ stability experiments. Each of the five vertical panels is for a distinct value of the leakage mobility parameter $L_m$.  Within each panel, the two values of the leakage parameter $L_1$ are represented with distinct colors. 
    The top plot of each panel shows the probability of logical error versus the number of rounds, and the lines are log-linear fits.  The solid line is for the simulations without wiggling, the dashed line is for the circuits with wiggling. At low mobility, the wiggled circuits perform better (the dashed curves are steeper), but at high mobility, non-wiggled circuits have higher accuracy (solid lines are steeper).  These are the results described in Fig.~\ref{fig:intersection_4x2}. 
    The bottom plot of each panel shows the probability of detecting at least one leaked qubit at each round of syndrome extraction of the 9-round experiment, which is the probability of extracting leakage per round.  Crosses correspond to wiggled circuits, dots to non-wiggled circuits.}
    \label{fig:results_4x2}
\end{figure}

In Fig.~\ref{fig:results_4x2}, we present additional data for the experiment described in the main text: the stability experiment for the $4\times 2$ rotated surface code patch of Fig.~\ref{fig:surface_code}. The top figure of each panel shows the data that goes into Fig.~\ref{fig:intersection_4x2}. All experiments are below threshold because the logical error decreases with increasing rounds, i.e.~$\Lambda_s>1$. The bottom panels show the amount of leakage observed at each round of syndrome extraction, which is equal to the amount of leakage removed at each round---since the measurement is followed by a reset.  At low mobility (left panels), wiggled circuits (crosses) remove more leakage than non-wiggled circuits (dots), showing that wiggling is a good LRU at low mobility.  However, at high mobility (right panels), we see that, on average, both wiggled (crosses) and non-wiggled (dots) circuits remove a similar amount of leakage per round. We also observe for the non-wiggled case that the amount of leakage removed at high mobility is higher than the amount removed at low mobility. This informs our interpretation of high leakage mobility as a leakage reduction unit.

There is an oscillating behavior between odd and even rounds in the amount of leakage removed by the wiggled circuits.  This is because leakage accumulates asymmetrically in alternating rounds, as qubits of different frequencies are involved in subsequent rounds of syndrome extraction (qubits at the center of dark/faint regions in Fig.~\ref{fig:surface_code}). Recall that, during a two-qubit gate, the higher-frequency member of the pair may leak, so rounds in which most qubits measured are higher-frequency extract more leakage.

\section{Simulation data for a $3\times 3$ unrotated surface code patch}
In this section, we describe an additional quantum stability experiment, involving the $3\times 3$ unrotated surface code patch.  The unrotated surface code is a different planar quantum error correction code that requires roughly twice as many qubits at a fixed distance.  We simulated this code as it has a larger ratio of bulk to boundary qubits compared to the rotated $4\times 2$, while still being simulable with the available resources. Having a larger bulk-to-boundary qubit ratio is more representative of larger patches, which will be needed for quantum utility.

\begin{figure}[h]
    \centering
    \includegraphics[width = 0.48\textwidth]{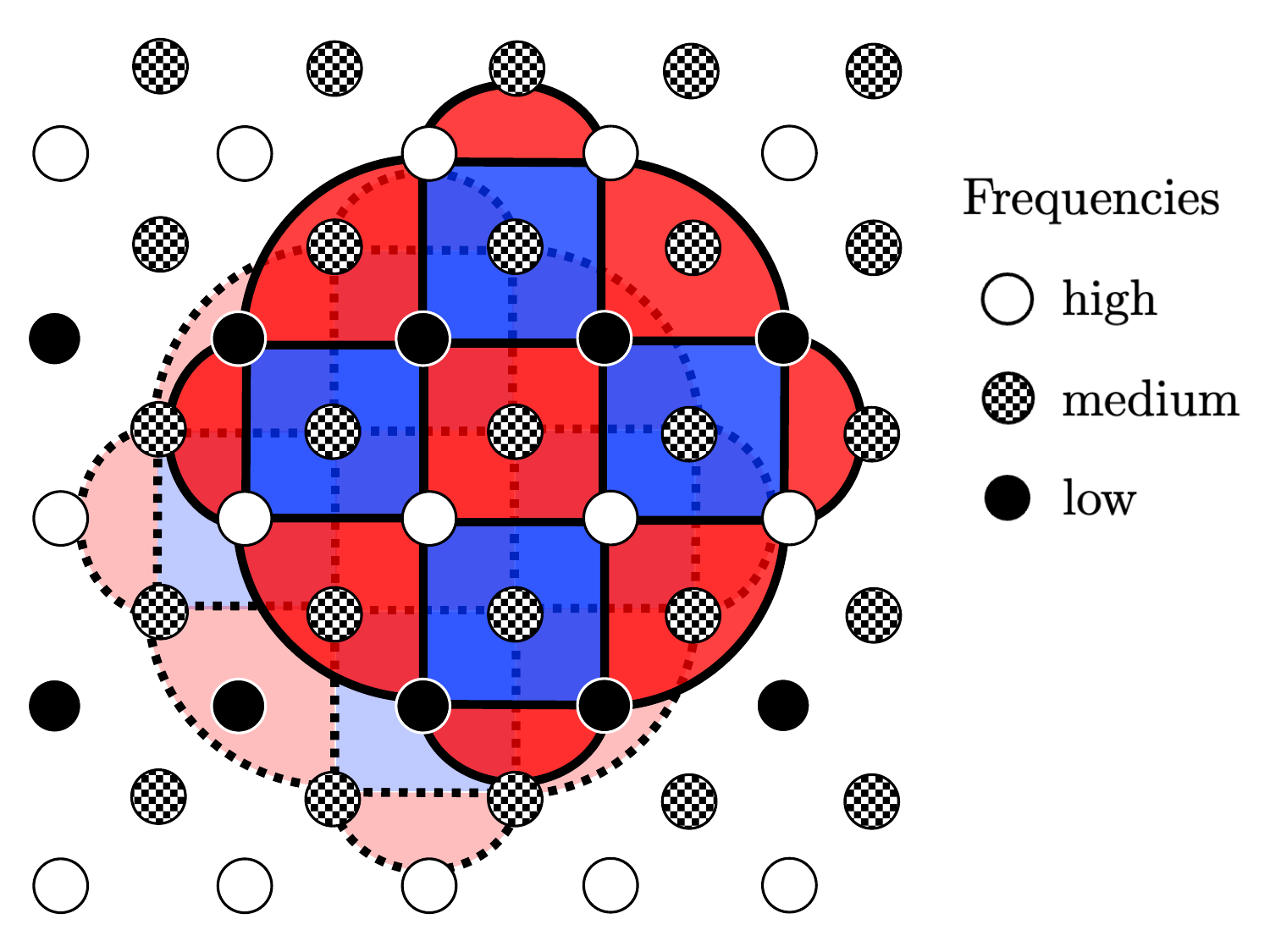}
    \caption{Wiggling surface code patch for the unrotated $3\times 3$ stability experiment. As in Fig.~\ref{fig:surface_code}, the location of the patch alternates between dark and faint in successive QEC rounds, and there are three types of qubits---high, medium, and low frequency.  In this experiment, the product of $X$ stabilizers (red) is the identity.}
    \label{fig:3x3_surface_code}
\end{figure}

\begin{figure*}[h]
    \centering
    \includegraphics[width = 0.48\textwidth]{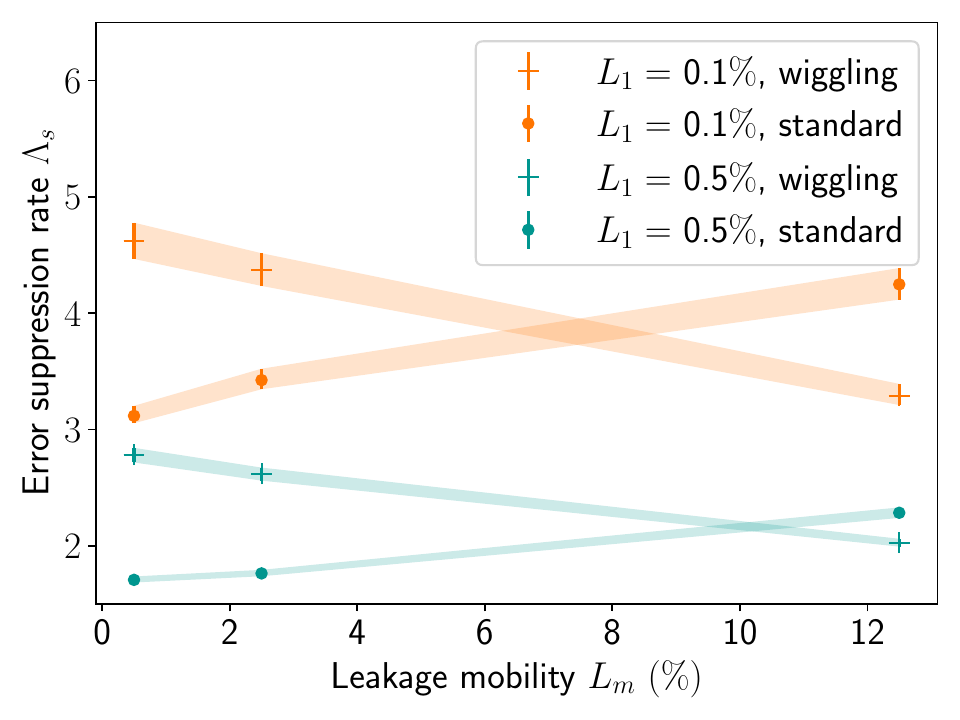}
    \caption{Results for the $3\times 3$ unrotated surface code stability experiment.  The results are qualitatively similar to the $4\times 2$ rotated surface code patch in Fig.~\ref{fig:intersection_4x2} in the main text.}
    \label{fig:intersection_3x3}
\end{figure*}

\begin{figure*}[h!]
    \centering
    \includegraphics[width = \textwidth]{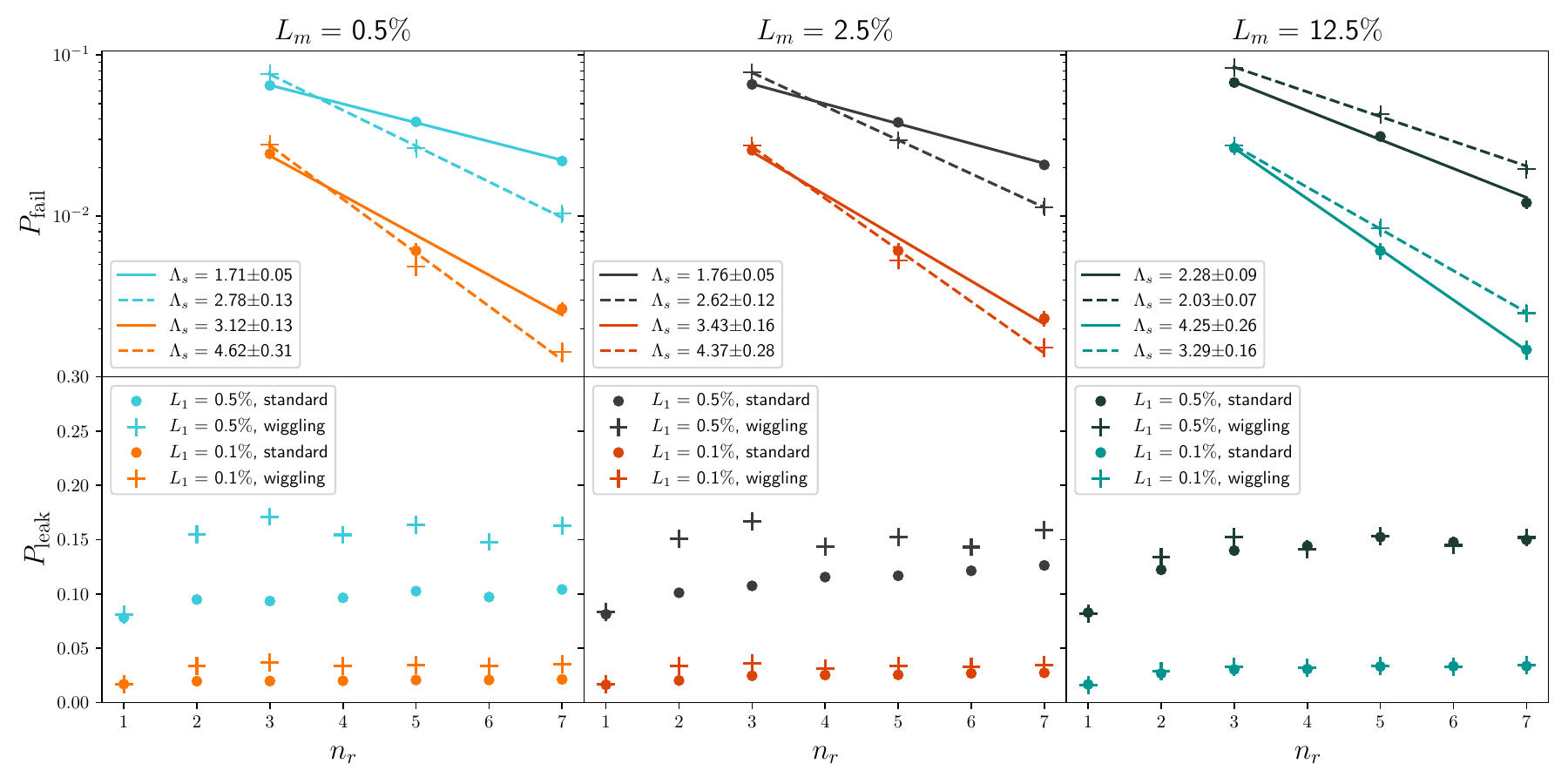}
    \caption{Results for the $3\times 3$ unrotated surface code stability experiment.  The results are qualitatively similar to the $4\times 2$ rotated experiment in Fig.~\ref{fig:results_4x2}.  Notice that the even/odd round oscillation in leakage extraction is smaller than in the $4\times 2$ rotated case, supporting the interpretation that this oscillation is a boundary effect.}
    \label{fig:results_3x3}
\end{figure*}

Fig.~\ref{fig:3x3_surface_code} describes the qubit patch and the frequency arrangement for the unrotated $3\times 3$ patch. Fig.~\ref{fig:intersection_3x3} and Fig.~\ref{fig:results_3x3} contain the results of our simulations, which are qualitatively similar to the simulations for the rotated $4\times 2$ patch: we observe a crossover between the performance of the wiggled and non-wiggled circuits at high mobility, and can interpret leakage mobility as a leakage reduction unit. Notice that the leakage extraction oscillation between odd and even rounds is less pronounced for unrotated $3\times 3$ (Fig.~\ref{fig:results_3x3}) compared to rotated $4\times 2$ (Fig.~\ref{fig:results_4x2})---evidencing that this oscillation is a boundary effect.

To explain this boundary effect, we consider the amount of leakage detected in a large square patch, where leakage at the boundary will be negligible compared to the bulk, in the absence of leakage mobility. In the bulk, at even rounds, half of the qubits we measure are high frequency. These qubits have leaked with probability $L_1$ times the number of CNOTs they have undergone since last measured (8 times). The other half are low-frequency qubits that do not leak. Therefore, averaging over all qubits at even rounds, we see leakage with probability $4L_1$. At odd rounds, we only measure medium-frequency qubits, which may leak half of the time they undergo a CNOT (when they interact with a low-frequency qubit)---4 times since last measured, totaling probability $4L_1$. Therefore, with large square patches, we would see the same amount of leakage at odd or even rounds.

\begin{thebibliography}{99}
\bibitem{nakamura_coherent_1999}
Y.~Nakamura, Y.~Pashkin, and J.~Tsai, Coherent control of macroscopic quantum states in a single-Cooper-pair box,
\href{https://doi.org/10.1038/19718}{Nature \textbf{398}, 786–788 (1999)}.

\bibitem{arute_quantum_2019}
F.~Arute, K.~Arya, R.~Babbush. \textit{et~al.} Quantum supremacy using a programmable superconducting processor,
\href{https://doi.org/10.1038/s41586-019-1666-5}{Nature \textbf{574}, 505–510 (2019)}.

\bibitem{wu_strong_2021}
Y.~Wu \textit{et~al.}, Strong Quantum Computational Advantage Using a Superconducting Quantum Processor,
\href{https://link.aps.org/doi/10.1103/PhysRevLett.127.180501}{Phys. Rev. Lett. \textbf{127}, 180501}.

\bibitem{kjaergaard_superconducting_2020}
M.~Kjaergaard, M.~E. Schwartz, J.~Braum\"uller, P.~Krantz, J.~I.-J. Wang, S.~Gustavsson, and W.~D. Oliver, Superconducting qubits: Current state of play, \href{https://doi.org/10.1146/annurev-conmatphys-031119-050605}{Annu. Rev. Condens. Matter Phys. \textbf{11}, 369 (2020)}.

\bibitem{bravyi_future_2022}
S.~Bravyi, O.~Dial, J.~M. Gambetta, D.~Gil, and Z.~Nazario, The future of quantum computing with superconducting qubits, \href{https://doi.org/10.1063/5.0082975}{J. Appl. Phys. \textbf{132}, 160902 (2022)}.

\bibitem{Li_error_2023}
Z.~Li, P.~Liu, P.~Zhao \textit{et~al.}, 
Error per single-qubit gate below 10-4 in a superconducting qubit,
\href{https://doi.org/10.1038/s41534-023-00781-x}{npj Quantum Inf \textbf{9}, 111 (2023)}.

\bibitem{aharonov_fault-tolerant_1999}
D.~Aharonov and M.~Ben-Or, Fault-tolerant quantum computation with constant error rate, \href{https://doi.org/10.48550/arXiv.quant-ph/9906129}{arXiv:quant-ph/9906129}.

\bibitem{suchara_leakage_2015}
M.~Suchara, A.~W. Cross, and J.~M. Gambetta, Leakage suppression in the toric code, in \href{https://doi.org/10.1109/ISIT.2015.7282629}{\textit{Proceedings of the 2015 IEEE International Symposium on Information Theory (ISIT)}, Hong Kong, 2015, p.~1119}.

\bibitem{miao_overcoming_2022}
K.~C. Miao \textit{et~al.}, Overcoming leakage in quantum error correction, \href{https://doi.org/10.1038/s41567-023-02226-w}{Nat. Phys. \textbf{19}, 1780 (2023)}.

\bibitem{acharya_suppressing_2023}
R.~Acharya \textit{et~al.} (Google Quantum AI), Suppressing quantum errors by scaling a surface code logical qubit, \href{https://doi.org/10.1038/s41586-022-05434-1}{Nature (London) \textbf{614}, 676 (2023)}.

\bibitem{wood_quantification_2018}
C.~J. Wood and J.~M. Gambetta, Quantification and characterization of leakage errors, \href{https://doi.org/10.1103/PhysRevA.97.032306}{Phys. Rev. A \textbf{97}, 032306 (2018)}.

\bibitem{ghosh_understanding_2013}
J.~Ghosh, A.~G. Fowler, J.~M. Martinis, and M.~R. Geller, Understanding the effects of leakage in superconducting quantum-error-detection circuits, \href{https://doi.org/10.1103/PhysRevA.88.062329}{Phys. Rev. A \textbf{88}, 062329 (2013)}.

\bibitem{fowler_coping_2013}
A.~G. Fowler, Coping with qubit leakage in topological codes, \href{https://doi.org/10.1103/PhysRevA.88.042308}{Phys. Rev. A \textbf{88}, 042308 (2013)}.

\bibitem{aliferis_fault-tolerant_2006}
P.~Aliferis and B.~M. Terhal, Fault-tolerant quantum computation for local leakage faults, \href{https://doi.org/10.48550/arXiv.quant-ph/0511065}{arXiv:quant-ph/0511065}.

\bibitem{mcewen_removing_2021}
M. McEwen \textit{et~al.}, Removing leakage-induced correlated errors in superconducting quantum error correction, \href{https://doi.org/10.1038/s41467-021-21982-y}{Nat. Commun. \textbf{12}, 1761 (2021)}.

\bibitem{battistel_hardware-efficient_2021}
F. Battistel, B.~M. Varbanov, and B.~M. Terhal, Hardware-efficient leakage-reduction scheme for quantum error correction with superconducting transmon qubits, \href{https://doi.org/10.1103/PRXQuantum.2.030314}{PRX Quantum \textbf{2}, 030314 (2021)}.

\bibitem{marques_all-microwave_2023-1}
J.~F. Marques \textit{et~al.}, All-microwave leakage reduction units for quantum error correction with superconducting transmon qubits, \href{https://doi.org/10.1103/PhysRevLett.130.250602}{Phys. Rev. Lett. \textbf{130}, 250602 (2023)}.

\bibitem{lacroix_fast_2023}
N.~Lacroix \textit{et~al.}, Fast flux-activated leakage reduction for superconducting quantum circuits, \href{https://doi.org/10.48550/arXiv.2309.07060}{arXiv:2309.07060}.

\bibitem{ghosh_leakage-resilient_2015}
J.~Ghosh and A.~G. Fowler, Leakage-resilient approach to fault-tolerant quantum computing with superconducting elements, \href{https://doi.org/10.1103/PhysRevA.91.020302}{Phys. Rev. A \textbf{91}, 020302 (2015)}.

\bibitem{brown_handling_2019}
N.~C. Brown, M.~Newman, and K.~R. Brown, Handling leakage with subsystem codes, \href{https://doi.org/10.1088/1367-2630/ab3372}{New J. Phys. \textbf{21}, 073055 (2019)}.

\bibitem{brown_leakage_2019}
N.~C. Brown and K.~R. Brown, Leakage mitigation for quantum error correction using a mixed qubit scheme, \href{https://doi.org/10.1103/PhysRevA.100.032325}{Phys. Rev. A \textbf{100}, 032325 (2019)}.

\bibitem{brown_critical_2020}
N.~C. Brown, A.~Cross, and K.~R. Brown, Critical faults of leakage errors on the surface code, in \href{https://doi.org/10.1109/QCE49297.2020.00043}{\textit{Proceedings of the 2020 IEEE International Conference on Quantum Computing and Engineering (QCE)}, Denver, 2020, p.~286}.

\bibitem{mcewen_relaxing_2023}
M.~McEwen, D.~Bacon, and C.~Gidney, Relaxing hardware requirements for surface code circuits using time-dynamics, \href{https://doi.org/10.22331/q-2023-11-07-1172}{Quantum \textbf{7}, 1172 (2023)}.

\bibitem{varbanov_leakage_2020}
B.~M. Varbanov, F.~Battistel, B.~M. Tarasinski, V.~P. Ostroukh, T.~E. O’Brien, L.~DiCarlo, and B.~M. Terhal, Leakage detection for a transmon-based surface code, \href{https://doi.org/10.1038/s41534-020-00330-w}{npj Quantum Inf. \textbf{6}, 102 (2020)}.

\bibitem{kitaev_fault-tolerant_2003}
A.~Yu. Kitaev, Fault-tolerant quantum computation by anyons, \href{https://doi.org/10.1016/S0003-4916(02)00018-0}{Ann. Phys. (Amsterdam) \textbf{303}, 2 (2003)}.

\bibitem{geher_error-corrected_2023}
G.~P. Geh\'er, C.~McLauchlan, E.~T. Campbell, A.~E. Moylett, and O.~Crawford, Error-corrected Hadamard gate simulated at the circuit level, \href{https://doi.org/10.48550/arXiv.2312.11605}{arXiv:2312.11605}.

\bibitem{gidney_stability_2022}
C.~Gidney, Stability experiments: The overlooked dual of memory experiments, \href{https://doi.org/10.22331/q-2022-08-24-786}{Quantum \textbf{6}, 786 (2022)}.

\bibitem{raussendorf_topological_2007}
R.~Raussendorf, J.~Harrington, and K.~Goyal, Topological fault-tolerance in cluster state quantum computation, \href{https://doi.org/10.1088/1367-2630/9/6/199}{New J. Phys. \textbf{9}, 199 (2007)}.

\bibitem{fowler_surface_2012}
A.~G. Fowler, M.~Mariantoni, J.~M. Martinis, and A.~N. Cleland, Surface codes: Towards practical large-scale quantum computation, \href{https://doi.org/10.1103/PhysRevA.86.032324}{Phys. Rev. A \textbf{86}, 032324 (2012)}.

\bibitem{dennis_topological_2002}
E.~Dennis, A.~Kitaev, A.~Landahl, and J.~Preskill, Topological quantum memory, \href{https://doi.org/10.1063/1.1499754}{J. Math. Phys. \textbf{43}, 4452 (2002)}.

\bibitem{higgott_pymatching_2022}
O.~Higgott, PyMatching: A Python package for decoding quantum codes with minimum-weight perfect matching, \href{https://doi.org/10.1145/3505637}{ACM Trans. Quantum Comput. \textbf{3}, 1 (2022)}.

\bibitem{delfosse_almost-linear_2021}
N.~Delfosse and N.~H. Nickerson, Almost-linear time decoding algorithm for topological codes, \href{https://doi.org/10.22331/q-2021-12-02-595}{Quantum \textbf{5}, 595 (2021)}.

\bibitem{horsman_surface_2012}
D.~Horsman, A.~G. Fowler, S.~Devitt, and R.~Van~Meter, Surface code quantum computing by lattice surgery, \href{https://doi.org/10.1088/1367-2630/14/12/123011}{New J. Phys. \textbf{14}, 123011 (2012)}.

\bibitem{chamberland_universal_2021}
C.~Chamberland and E.~T. Campbell, Universal quantum computing with twist-free and temporally encoded lattice surgery, \href{
https://doi.org/10.1103/PRXQuantum.3.010331}{PRX Quantum \textbf{3}, 010331 (2022)}.

\bibitem{chamberland_circuit-level_2022}
C.~Chamberland and E.~T. Campbell, Circuit-level protocol and analysis for twist-based lattice surgery, \href{https://doi.org/10.1103/PhysRevResearch.4.023090}{Phys. Rev. Research \textbf{4}, 023090 (2022)}.

\bibitem{geher_tangling_2023}
G.~P. Geh\'er, O.~Crawford, and E.~T. Campbell, Tangling schedules eases hardware connectivity requirements for quantum error correction, \href{https://doi.org/10.1103/PRXQuantum.5.010348}{PRX Quantum \textbf{5}, 010348 (2024)}.

\bibitem{krinner_realizing_2022}
S.~Krinner \textit{et~al.}, Realizing repeated quantum error correction in a distance-three surface code, \href{https://doi.org/10.1038/s41586-022-04566-8}{Nature (London) \textbf{605}, 669 (2022)}.

\bibitem{ali_reducing_2024}
H.~Ali, J.~Marques, O.~Crawford, J.~Majaniemi, M.~Serra-Peralta, D.~Byfield, B.~Varbanov, B.~M. Terhal, L.~DiCarlo, and E.~T. Campbell, Reducing the error rate of a superconducting logical qubit using analog readout information, \href{https://doi.org/10.48550/arXiv.2403.00706}{arXiv:2403.00706}.

\bibitem{bravyi_correcting_2018}
S.~Bravyi, M.~Englbrecht, R.~König and N.~Peard, Correcting coherent errors with surface codes,
\href{https://doi.org/10.1038/s41534-018-0106-y}{npj Quantum Inf 4, 55 (2018)}. 

\bibitem{katsuda_simulation_2022}
M.~Katsuda, K.~Mitarai, and K.~Fujii, Simulation and performance analysis of quantum error correction with a rotated surface code under a realistic noise model, \href{https://doi.org/10.1103/PhysRevResearch.6.013024}{Phys. Rev. Research \textbf{6}, 013024 (2024)}.

\bibitem{manabe_efficient_2023}
H.~Manabe, Hidetaka, Y.~Suzuki and A.~S.~Darmawan, Efficient Simulation of Leakage Errors in Quantum Error Correcting Codes Using Tensor Network Methods,
\href{https://doi.org/10.48550/arXiv.2308.08186}{arXiv:2308.08186}

\bibitem{bravyi_trading_2016}
S.~Bravyi, G.~Smith, and J.~A. Smolin, Trading classical and quantum computational resources, \href{https://doi.org/10.1103/PhysRevX.6.021043}{Phys. Rev. X \textbf{6}, 021043 (2016)}.

\bibitem{suzuki_Qulacs_2021}
Y.~Suzuki \textit{et~al.}, Qulacs: A fast and versatile quantum circuit simulator for research purpose, \href{https://doi.org/10.22331/q-2021-10-06-559}{Quantum \textbf{5}, 559 (2021)}.

\bibitem{imamura_mpiQulacs_2022}
S.~Imamura, M.~Yamazaki, T.~Honda, A.~Kasagi, A.~Tabuchi, H.~Nakao, N.~Fukumoto, and K.~Nakashima, mpiQulacs: A distributed quantum computer simulator for A64FX-based cluster systems, \href{https://doi.org/10.48550/arXiv.2203.16044}{arXiv:2203.16044}.

\bibitem{krantz_guide_2019}
P.~Krantz, M.~Kjaergaard, F.~Yan, T.~P. Orlando, S.~Gustavsson, and W.~D. Oliver, A quantum engineer's guide to superconducting qubits, \href{https://doi.org/10.1063/1.5089550}{Appl. Phys. Rev. \textbf{6}, 021318 (2019)}.

\bibitem{blais_circuit_2021}
A.~Blais, A.~L. Grimsmo, S.~M. Girvin, and A.~Wallraff, Circuit quantum electrodynamics, \href{https://doi.org/10.1103/RevModPhys.93.025005}{Rev. Mod. Phys. \textbf{93}, 025005 (2021)}.

\bibitem{versluis_scalable_2017}
R.~Versluis, S.~Poletto, N.~Khammassi, B.~Tarasinski, N.~Haider, D.~J. Michalak, A.~Bruno, K.~Bertels, and L.~DiCarlo, Scalable quantum circuit and control for a superconducting surface code, \href{https://doi.org/10.1103/PhysRevApplied.8.034021}{Phys. Rev. Applied \textbf{8}, 034021 (2017)}.

\bibitem{lindblad_generators_1976}
G.~Lindblad, On the generators of quantum dynamical semigroups,
\href{https://doi.org/10.1007/BF01608499}{Commun. Math. Phys. 48, 119–130 (1976)}.

\bibitem{rol_fast_2019}
M.~A. Rol \textit{et~al.}, Fast, high-fidelity conditional-phase gate exploiting leakage interference in weakly anharmonic superconducting qubits, \href{https://doi.org/10.1103/PhysRevLett.123.120502}{Phys. Rev. Lett. \textbf{123}, 120502 (2019)}.

\bibitem{obrien_density-matrix_2017}
T.~E. O’Brien, B.~Tarasinski, and L.~DiCarlo, Density-matrix simulation of small surface codes under current and projected experimental noise, \href{https://doi.org/10.1038/s41534-017-0039-x}{npj Quantum Inf. \textbf{3}, 39 (2017)}.

\bibitem{spitz_adaptive_2018}
S.~T. Spitz, B.~Tarasinski, C.~W.~J. Beenakker, and T.~E. O’Brien, Adaptive weight estimator for quantum error correction in a time-dependent environment, \href{https://doi.org/10.1002/qute.201800012}{Adv. Quantum Technol. \textbf{1}, 1800012 (2018)}.

\bibitem{chen_calibrated_2022}
E.~H. Chen, T.~J. Yoder, Y.~Kim, N.~Sundaresan, S.~Srinivasan, M.~Li, A.~D. C\'orcoles, A.~W. Cross, and M.~Takita, Calibrated decoders for experimental quantum error correction, \href{https://doi.org/10.1103/PhysRevLett.128.110504}{Phys. Rev. Lett. \textbf{128}, 110504 (2022)}.

\bibitem{chen_exponential_2021}
Z.~Chen \textit{et~al.} (Google Quantum AI), Exponential suppression of bit or phase errors with cyclic error correction, \href{https://doi.org/10.1038/s41586-021-03588-y}{Nature (London) \textbf{595}, 383 (2021)}.

\bibitem{gidney_stim_2021}
C.~Gidney, Stim: A fast stabilizer circuit simulator, \href{https://doi.org/10.22331/q-2021-07-06-497}{Quantum \textbf{5}, 497 (2021)}.

\bibitem{nielsen_quantum_2010}
M.~A. Nielsen and I.~L. Chuang, \textit{Quantum Computation and Quantum Information} (Cambridge University Press, Cambridge, UK, 2010).
\end{thebibliography}
\end{document}